\documentclass{article}
\usepackage{spconf,amsmath}
\usepackage{graphicx}
\usepackage{listings}
\usepackage{algorithm}
\usepackage{algcompatible}
\usepackage{algpseudocode}

\usepackage{amsmath,amssymb}
\usepackage{hyperref}
\usepackage{cleveref}
\usepackage{subcaption}
\usepackage{stmaryrd}
\usepackage{booktabs}
\usepackage{tabularx}
\usepackage{dirtytalk}
\usepackage{multirow}
\usepackage{adjustbox}
\usepackage{comment}
\usepackage{bm}
\usepackage{xcolor}

\usepackage[acronym]{glossaries}
\newacronym{us}{US}{Ultrasound}
\newacronym{fps}{fps}{frames per second}
\newacronym{ef}{LVEF}{Left Ventricular Ejection Fraction}
\newacronym{es}{ES}{End-Systolic}
\newacronym{ed}{ED}{End-Diastolic}
\newacronym{sv}{SV}{Systolic Volume}
\newacronym{cdm}{CDM}{Cascaded Diffusion Model}
\newacronym{edm}{EDM}{Elucidated Diffusion Model}
\newacronym{ml}{ML}{Machine Learning}
\newacronym{mri}{MRI}{Magnetic Resonance Imaging}
\newacronym{ct}{CT}{Computed Tomography}

\newcommand{\etal}{\mbox{\emph{et al.}}}
\usepackage{caption}
\makeatletter
\newcommand*{\inlineequation}[2][]{%
  \begingroup
    \refstepcounter{equation}%
    \ifx\\#1\\%
    \else
      \label{#1}%
    \fi
    \relpenalty=10000 %
    \binoppenalty=10000 %
    \ensuremath{%
      #2%
    }%
    ~\@eqnnum
  \endgroup
}
\makeatother

\title{Echocardiography video synthesis from end diastolic semantic map via diffusion model}

\name{Nguyen Van Phi$^{1}$, Tran Minh Duc$^{1}$, Pham Huy Hieu$^{2\dag}$, Tran Quoc Long $^{1\dag}$ \thanks{$\dag$ Corresponding author.} \thanks{Supported by VINIF under project code VINIF.2019.DA02.}}
\address{ $^1$ University of Engineering and Technology, VNU, Hanoi, Vietnam\\
$^2$VinUni-Illinois Smart Health Center, VinUni, Hanoi, Vietnam}
\begin{document}
%


\maketitle

\begin{abstract}
Denoising Diffusion Probabilistic Models (DDPMs) have demonstrated significant achievements in various image and video generation tasks, including the domain of medical imaging. However, generating echocardiography videos based on semantic anatomical information remains an unexplored area of research. This is mostly due to the constraints imposed by the currently available datasets, which lack sufficient scale and comprehensive frame-wise annotations for every cardiac cycle. This paper aims to tackle the aforementioned challenges by expanding upon existing video diffusion models for the purpose of cardiac video synthesis. More specifically, our focus lies in generating video using semantic maps of the initial frame during the cardiac cycle, commonly referred to as end diastole. To further improve the synthesis process, we integrate spatial adaptive normalization into multiscale feature maps. This enables the inclusion of semantic guidance during synthesis, resulting in enhanced realism and coherence of the resultant video sequences. Experiments are conducted on the CAMUS dataset, which is a highly used dataset in the field of echocardiography. Our model exhibits better performance compared to the standard diffusion technique in terms of multiple metrics, including FID, FVD, and SSMI.

\end{abstract}

\begin{figure}[hbt]
    \centering
    \includegraphics[width=\linewidth]{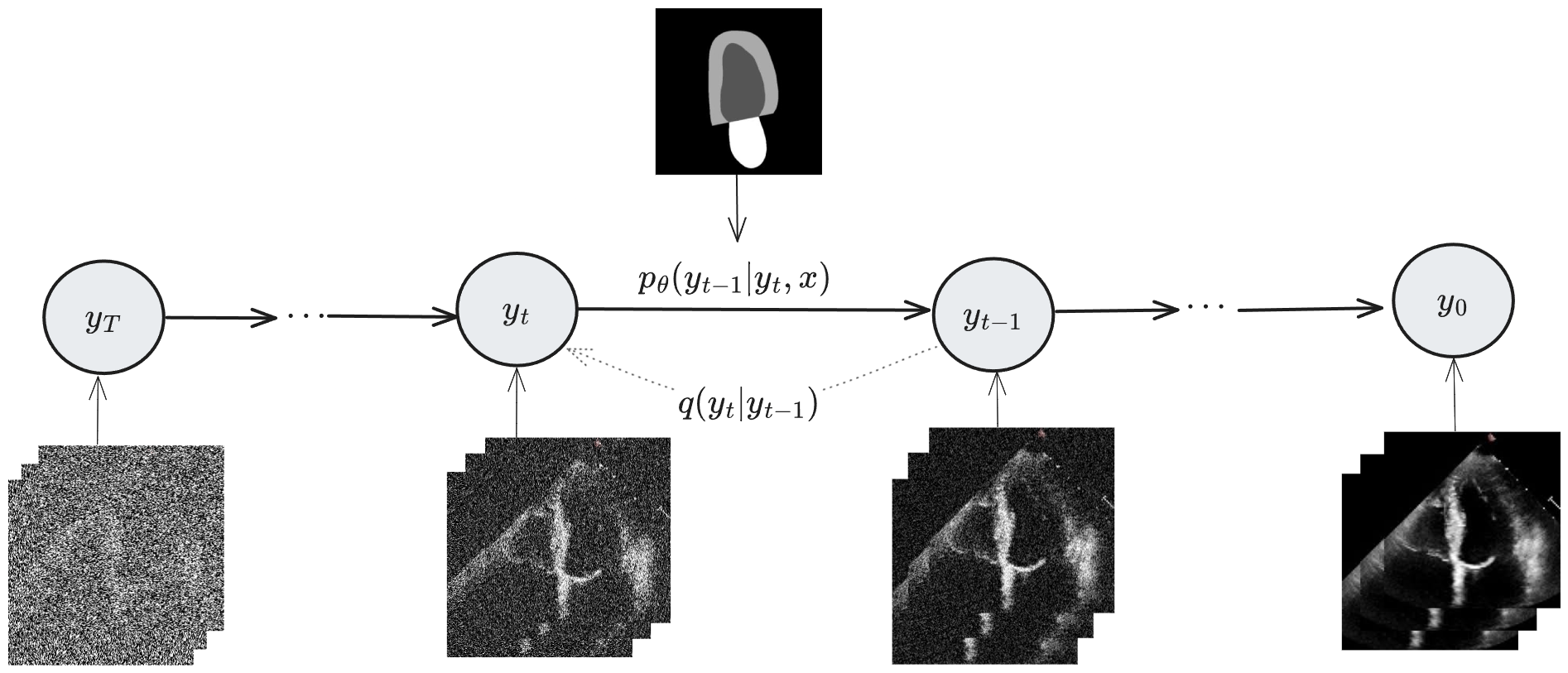}
    \setlength{\belowcaptionskip}{-10pt} 
    \caption{\textbf{Conditional Diffusion Model for Semantic Echocardiography Video Synthesis.} Our framework transforms a tensor of standard Gaussian noise into a realistic echocardiography video via iterative denoising process, given the guidance of the semantic label map $x$}
    \label{fig:ddpms}
\end{figure}

\section{Introduction}
Echocardiography is an ultrasound of the heart that supports cardiologists in assessing the structure and function of the heart. It is widely used in diagnosis thanks to its accessibility, affordability, and non-invasiveness. In recent years, many research efforts have been made to apply machine learning to echocardiography to improve image analysis, automate diagnostic tasks, and advance our understanding of cardiac conditions \cite{zhou2021artificial,ghorbani2020deep}.
However, one of the biggest challenges in echocardiography today is the degradation of image quality caused by the ultrasound image formation process. Ultrasound images frequently contain speckle noise and motion artifacts, which can result in inaccurate anatomical structure examination and expensive manual annotation of echocardiograms.~\cite{varoquaux2022machine}.
The traditional method of developing machine learning involves collecting and manually labeling a large number of data samples. Manual annotations must be done by clinical specialists, which is costly and time-consuming. Synthesizing ultrasound images has recently come to light as a promising method for obtaining a wide range of reliable datasets for training machine learning models \cite{Gilbert2021}.

Two primary approaches are frequently utilized for synthesizing ultrasound images: physics-based simulation and machine learning-based image generation.
The goal of physics-based simulation is to reproduce the processes of beamforming and ultrasound formation, seeking to mimic their behavior and characteristics~\cite{burger_real-time_2013,garcia2022simus}.
However, to accurately simulate the underlying physics, these methods require realistic scatter maps of the heart.
Moreover, obtaining these realistic scatter maps at a large scale is extremely difficult.

On the other hand, conditional ultrasound image generation has demonstrated promising outcomes ~\cite{Gilbert2021,reynaud2023featureconditioned,stojanovski2023echo}.
Previously, Generative Adversarial Networks (GANs) have been the go-to solution to generate ultrasound sequences~\cite{liang2022weakly}.
Multiple approaches were also put forth to control the model's generating behavior. Liang \etal~\cite{liang_sketch_2022} proposed a sketch guided GANs to obtain an editable image synthesis model.
Gilbert \etal~\cite{Gilbert2021} suggested using GANs conditioned on semantic maps from 3D deformable cardiac models to generate ultrasound images. Nevertheless, GANs still suffered from poor mode coverage~\cite{zhao2018bias}.

Recently, diffusion models~\cite{ho_denoising_2020} have emerged as powerful generative models and demonstrated their effectiveness in generating realistic data.
They could produce ultrasound images and videos based on conditions such as clinical attributes or segmentation maps.
For example, Reynaud \etal~\cite{reynaud2023featureconditioned} utilized cascaded diffusion model conditioned on End Diastolic (ED) frame to generate ultrasound image with various left ventricle ejection fractions (LVEFs) levels. However, the model requires a real ultrasound image as an initial condition, limiting the diversity of generated images.
Stojanovski \etal~\cite{stojanovski2023echo} also used diffusion models but use semantic label maps as conditions. But their model was designed to generate single image, and extending it to generate image sequences is non-trivial due to the unavailability of fully annotated ultrasound videos.
A solution is to convert the 2D Convolutional Neural Networks (CNNs) into 3D CNNs~\cite{wang2018video}. However, this approach is computationally inefficient and requires a lot of memory. Pan \etal~\cite{pan2019video} proposed a method that generates videos based on optical flow, but accurate optical flow estimation is necessary for its success.
As a result, little has been done to synthesize echocardiography videos using single semantic label map. 
This presents a unique challenge that needs to be addressed to be able to generate coherent and accurate echocardiography sequences.

\noindent\textbf{Contribution.} 
In this paper, we present an echocardiography video generation model based on DDPMs~\cite{ho_denoising_2020,ho_video_2022}. By dynamically incorporating the semantic label map of diastolic in multi-scale decoder, our model could produce realistic echocardiography sequences with diverse anatomical structures.
Our contributions could be summarized as follows: (1) To the best of our knowledge, this study is the first attempt at generating echocardiography video from a semantic label map using diffusion model. (2) We propose a new network structure to handle noisy input and semantic mask effectively in order to incorporate anatomical information and produce realistic ultrasound sequences.

\section{Method}

\begin{figure}[hbt]
    \centering
    \includegraphics[width=\linewidth]{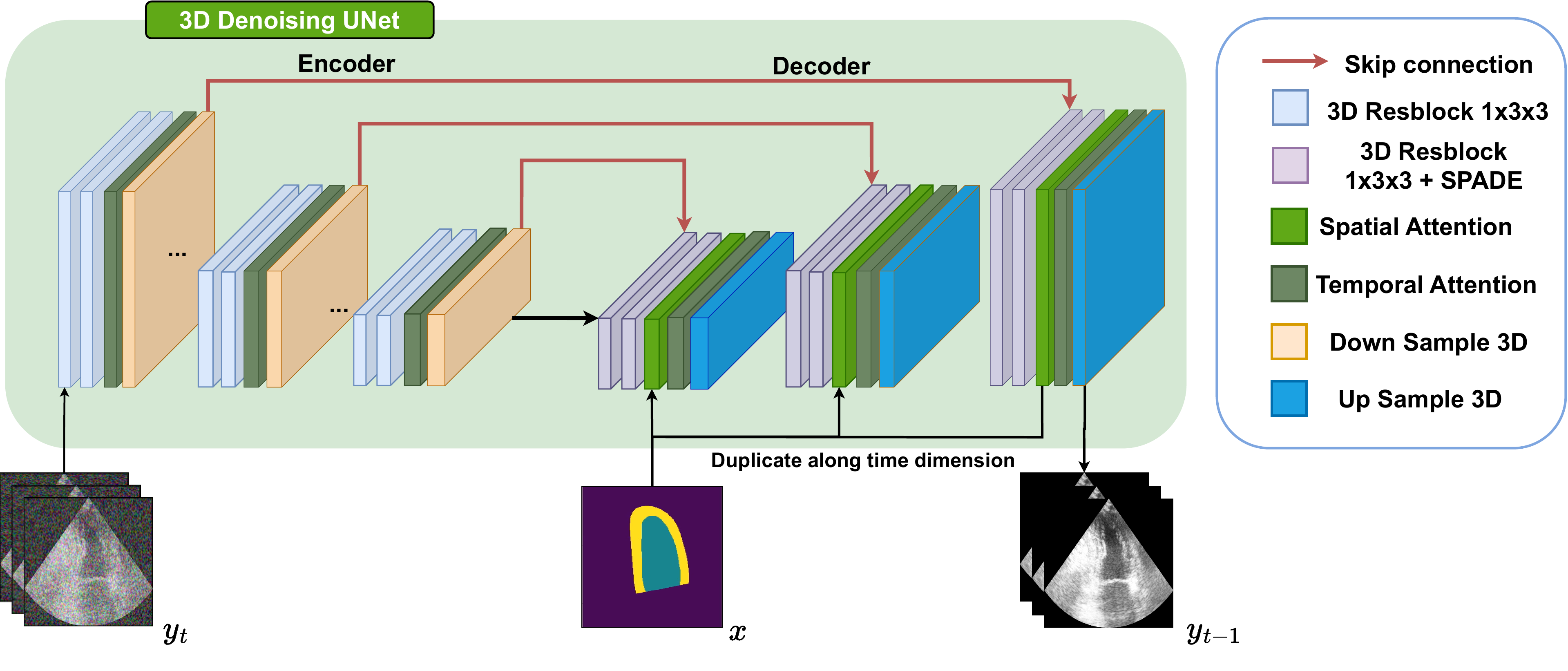}
    \caption{Our proposed network architecture. The 3D Denoising UNet takes the semantic segmentation map of the ED frame $x$ and the perturbed $y_t$ at denoising step $t$. The model predict the noise $\epsilon_\theta$ for reverse process to obtain synthesis sample $\hat{y}_i$ corresponding to $x$.}
    \label{fig:narrow_model}
\end{figure}

Our model generates a new video by gradually removing Gaussian noise under the guidance from semantic segmentation map (see Figure~\ref{fig:ddpms}). In the next section, we will describe the conditional DDPMs and explain how we incorporate semantic information into the denoising process.

\noindent
\textbf{Conditional DDPMs.} There are two Markov process involved in DDPMS, i.e. forward process and reverse process.
The forward process progressively adds noise into data, whereas the reverse process tries to eliminate it. Given the condition $x$,
the goal of conditional DDPMs is to maximize the likelihood $p_\theta (y_0|x)$ while the conditional data follows to a distribution $q(y_0|x)$.
Starting from a Gaussian noise $p(y_T) \sim \mathcal{N}(0, \mathbf{I})$, the reverse process $p_\theta (y_{0:T}|x)$ is a Markov process with learned Gaussian transitions, which is formulated as follows:
\begin{equation}
    p_\theta (y_{0:T}|x) = p(y_T)\prod_{t=1}^T p_\theta (y_{t-1}|y_t, x)
\end{equation}
\begin{equation}
    p_\theta (y_{t-1}|y_t, x) = \mathcal{N}(y_{t-1}; \mu_\theta(y_t, x, t), \Sigma_\theta(y_t, x, t))
\end{equation}
\noindent
The forward process takes a data sampled from a real data distribution $q(y_0)$ and iteratively perturbing the data by adding a small Gaussian noise according to a variance schedule $\beta_1,\dots \beta_T$. The transition distribution is formulated as follows:
\begin{equation}
    q(y_t|y_{t-1}) = \mathcal{N}(y_t; \sqrt{1-\beta_t}y_{t-1}, \beta_t\mathbf{I})
\end{equation}
Let $\alpha_t = \prod_{s=1}^t(1-\beta_t)$, we could compute the transition distribution of $y_t$ given $y_0$ directly with formula:
\begin{equation}
    q(y_t|y_0) = \mathcal{N}(y_t; \sqrt{\alpha_t}y_0, (1-\alpha_t)\mathbf{I})
\end{equation}
The conditional DDPMs is trained by maximizing the Evidence Lower Bound. By applying reparameterization trick, this is equivalent to minimizing the discrepancy between the noise added in the forward process and the noise removed during the reverse process. Therefore, the objective function at time step $t$ is defined as follow:
\begin{equation}
    \mathcal{L}_{t} = \mathbb{E}_{y_0 \sim q(y_0), \epsilon \sim \mathcal{N}(0, \mathbf{I})} {\parallel \epsilon - \epsilon_{\theta}(y_t, x, t)\parallel}^2
\end{equation}
\noindent
where $t$ is sampled uniformly from the range $[1\dots T]$ and $y_0$ is sampled from real data distribution $q(y_0)$. In the context of our study, $y_0$ is a sequence of frames captured within a cardiac cycle $y_0 \in \mathbb{R}^{K\times C\times H\times W}$. Where $K$ represents the fixed number of selected frames in one video and $C, H, W$ is the spatial dimensions of each frame. Moreover, each cycle was given an annotated semantic map of the first frame $x \in \mathbb{R}^{C\times H\times W}$, which will serve as the condition for our model. This semantic map has same spatial dimension as the original images. Thus, our goal is to learn a model that could generate realistic data from given semantic structure.

\noindent
\textbf{Semantic Conditioned Diffusion Model.} Figure~\ref{fig:narrow_model} shows an overview of our conditional denoising network architecture, which is based on 3D-Unet proposed by Ho \etal~\cite{ho_video_2022}. 
The denoising encoder receives the noisy image sequence and computes the feature representations. The decoder then uses these feature vectors and the injected semantic information to reconstruct the real images.

Since our input is a sequence of frames, we used a stack of multiple 3D Residual Convolution Blocks as our encoder. For each block, 3D Convolution layers were used to compute the feature representations. The time step information $t$ was encoded by cosine embedding and then added to every feature outputs. Group normalization is then used to normalize those features. Further more,
we used a spatial attention layer followed by a temporal attention layer in each block to allow the model to learn the spatial and temporal relationships between each frame.

In the decoder, each residual block was modified so that the condition information, which is the semantic map describing the structure of the heart, could be effectively injected.
Saharia \etal \cite{saharia2022palette} showed that directly concatenate the condition information and noisy images as input does not fully leverage the semantic information. Whereas, Wang \etal \cite{wang2022semantic} demonstrated the effective of Spatial Adaptive Normalization (SPADE) for adding the semantic label map. The features were regulated by the SPADE in a learnable, spatially-adaptive manner. Therefore, we inject the semantic label map using SPADE layer over Group Normalization layer.
Specifically, given a feature vector $f_i$ of input images from a decoder block, we want to add the condition information $x$, which is the semantic label map of the first frame. 
Since $x$ does not initially match the size of the input images, it must be duplicated along the temporal axis.
The normalization is formulated as follows:
\begin{equation} f^{i+1} = \gamma^i(x, k) \cdot \text{Norm}(f^i) + \delta^i(x, k)\end{equation}
where $f^i$ and $f^{i+1}$ are the input and output features of SPADE. $\text{Norm}(\cdot)$ is parameter-free group normalization.
The $\gamma^i(x, k)$ and $\delta^i(x, k)$ are the spatially-adaptive weight and bias learned from the semantic label map $x$ and cosine embedding $k$ of frame time. Since we only inject the label map of the first frame, $k$ was added to provide the temporal information.

Inspired by \cite{ho_classifier-free_2022}, we applied classifier free approach to train our model. Since it is showed that the gradient $\nabla_{y} \text{log} p(x|y)$ of an extra classifier could improve samples from conditional diffusion models \cite{dhariwal2021diffusion}.
The key idea is to replace the semantic label map $x$ with a null label $\emptyset$ under certain probability. \cite{ho_classifier-free_2022} showed that this technique implicitly infers the gradient of the log probability. The sampling procedure is obtained as following formula:
\begin{equation}\epsilon_\theta(y_t|x) = \epsilon_\theta(y_t|x) + s \cdot (\epsilon_\theta(y_t|x) - \epsilon_\theta(y_t|\emptyset))\end{equation}
In our implementation, $\emptyset$ is a black image with all-zero elements.


\section{Experiments}
\noindent\textbf{Dataset.}
Our experiments were conducted on the CAMUS dataset~\cite{leclerc2019deep}.
There are 450 patients in this dataset, and each has three recorded chamber views. 
For simplicity, we only conducted experiments on 2 chamber view videos in our study.
For each data sample, we have a video of a complete cardiac cycle, from the ED phase to ES phase.
However, only the ED and ES frames have semantic map annotation, which were labeled by cardiologists.
There are four classes on each segmentation map: background, epicardium, myocardium, and left atrium.
To avoid data leakage, we split the dataset by patients using 80-10-10 ratio. As a result, the training set contains 360 patients, the validation and test set contains 45 patients, respectively.

\noindent\textbf{Baselines.}
Since our model was based on DDPMs, we used this model as the baseline. Besides, we also implemented the cascade diffusion architecture~\cite{ho_cascaded_2022} to validate the efficacy of this technique for semantic conditional generation. Since this model have been shown to generate videos more efficiently, including unconditioned ultrasound generation~\cite{reynaud2023featureconditioned}.
In addition to these two primary architectures, we have implemented SPADE and concatenation as two condition features injection approaches.
We validated our models using variety of number frames settings, including taking 16 or 24 frames.

\noindent\textbf{Experimental Settings.}
Every models were trained on a node with three NVIDIA A100 gpus.
We set the batch size of 24 for three GPUs.
We chose the total diffusion steps $T=1000$, and the classifier-free guidance factor $s=7.0$ was used. We used Adam optimizer with learning rate $lr=1e-4$ in every training.
Two UNet backbones were utilized for the cascade network architecture. One for low-resolution video synthesis, i.e. generate sequences with dimension of $n_{frame} \times 56\times 56$. The second one is for super-resolution which converts output with spatial dimension of $56 \times 56$ into $128 \times 128$.
We kept all of the settings the same to ensure a fair comparision.

\noindent\textbf{Evaluation Metrics.}
We assessed the models' performances using three metrics. Which are the Fréchet Inception Distance (FID)~\cite{heusel2017gans}, Fréchet Video Distance (FVD)~\cite{unterthiner2018towards}, and Structure Similarity Index (SSIM) score~\cite{ssim}.
FID and FVD have been commonly used in many studies to measure generated images and videos quality.
FID computes the distance between two distributions, one from generated images and the other from real images, whereas FVD does the same for videos. 
A lower score indicates higher quality in terms of visual fidelity, diversity, and temporal consistency of the generated videos.
SSIM score is used for measuring the similarity between two images. In our study, we calculate this score by averaging SSIM between frames of generated videos with frames from ground truth videos, while both have the same segmentation map.
Higher SSIM score indicates higher similarity between synthetic frames and the ground-truth frames.
We generated totally  450 videos for the test set, with 10 videos for each segmentation map.

\begin{table}[hbt]
\begin{tabular}{llccccc}
\toprule
                         
Cond. & Model & $K$ & FID$\downarrow$  & FVD$\downarrow$ & SSIM$\uparrow$ \\
\hline
\multirow{4}{*}{Concat}  
                         & \multirow{2}{*}{Cascade}      & 16  & 42.93 & 278.35 & 0.47  \\ \cmidrule(r){3-6} 
                         &                               & 24 & 50.93 & 310.05 & 0.51   \\ \cmidrule(r){2-6} 
                         & \multirow{2}{*}{DDPM}         & 16  & 28.56  & 137.73 & 0.55 \\ \cmidrule(r){3-6} 
                         &                               & 24  & 21.46 & 144.79 & 0.53  \\ \hline \\
\multirow{4}{*}{SPADE(our)} 
                         & \multirow{2}{*}{Cascade}      & 16 & 34.52 & 214.45 & 0.49   \\ \cmidrule(r){3-6} 
                         &                               & 24 & 40.87 & 231.88 & 0.52   \\ \cmidrule(r){2-6} 
                         & \multirow{2}{*}{DDPM}         & 16  & 18.65 & \textbf{89.78}  & \textbf{0.56}  \\ \cmidrule(r){3-6} 
                         &                               & 24  & \textbf{16.05} & 115.79 & 0.54  \\ \hline
\end{tabular}
\caption{Quantitative comparison with existing methods on semantic echocardiography video synthesis. $\uparrow$ indicates the higher the better, while $\downarrow$ indicates the lower the better. Notably, our method achieves state-of-the-art performance on all metrics.}
\label{tab:res}
\end{table}

\noindent\textbf{Results.}
Table~\ref{tab:res} presents the results of the methods on various metrics.
Overall, using SPADE as the input to the denoising model instead of concatenating the segmentation map and ultrasound led to an improvement in the quality of ultrasound images from both a single image and a video perspective.
For instance, in the case of DDPM, SPADE has FID of 16.05 as compared to concatenation's 21.46 and FVD of 115.79 as compared to concatenation's 144.79.
When comparing cascade and DDPM, we noticed that DDPM produce better performance, with FVD of 89.78 for DDPM versus 214.45 for cascade.
One explanation could be that the DDPM architecture uses convolution layers and attention layers for the entire input resolution, but the cascade model only performs temporal attention on downsampled versions of inputs.
The cascade approach, however, is more effective in terms of processing and memory.
Additionally, we found that the amount of frames had no noticeable effect on the model's performance.
This could be as a result of the segmentation map being the same for all video frames.
Finally, SSIM score are generally better when using SPADE than those using concatination.

\begin{figure}[hbt]
    \center
    \includegraphics[width=\linewidth]{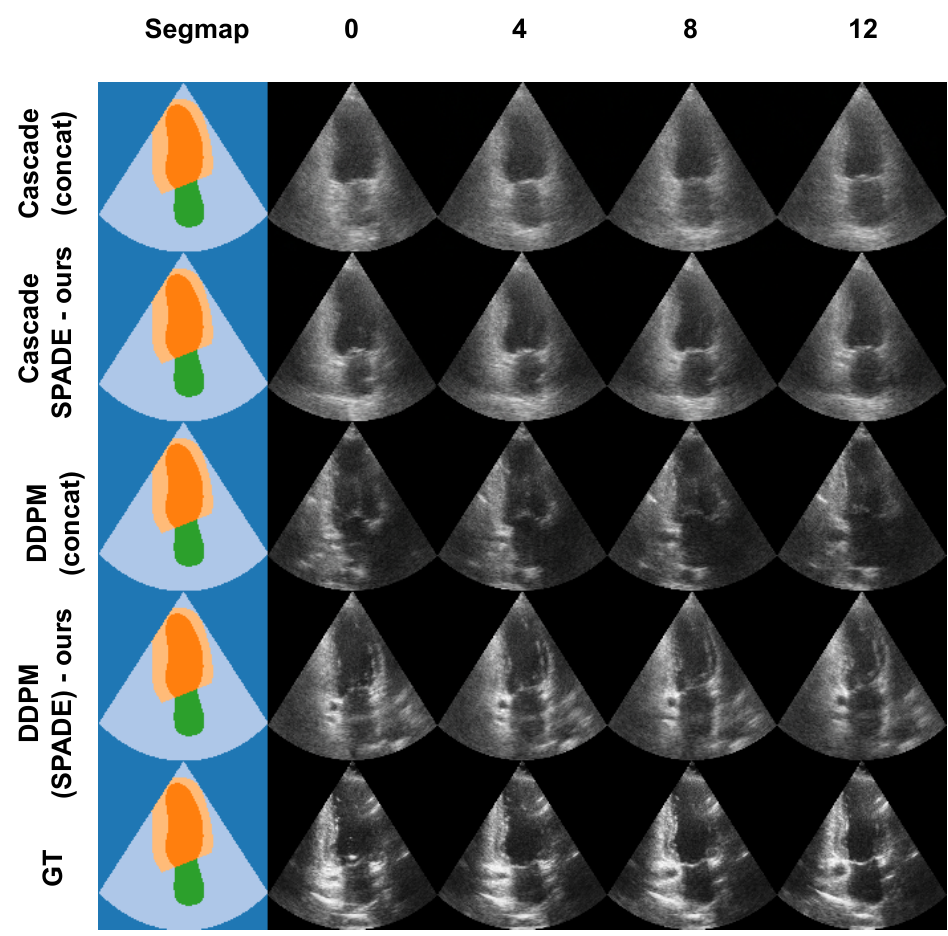}
    \caption{Qualitative results on the CAMUS dataset. All models were conditioned on the same segmentation map of the ED frame. We selected frames every 4 time steps to show the temporal change in one video. More videos can be found at \url{https://tinyurl.com/5n8m6k92}.}
    \label{fig:example}
\end{figure}


A visual comparison of different approaches is shown in Figure~\ref{fig:example}.
In general, the images produced by our suggested method using SPADE have higher fidelity and closely resemble actual ultrasound images. 
More specifically, our model generates images with sharper edges and more realistic anatomical structures, especially in the region of endocardium and myocardium. 
While the original DDPMs with concatination produces images with blurry edges and artifacts, SPADE could produce images with perceivable speckle motion.
Comparing DDPM and Cascade, we found that DDPM produces better images in terms of visual fidelity and speckle motion over time. While our photos and the SPADE's ground truth have extremely similar anatomical structures, there are still some artifacts and blurry speckles in them. We also recommended that human review should be done to better understand the quality of the generated videos and that future work should take into account training the models on longer, higher resolution videos that show different moments in the cardiac cycle.

\section{Conclusion}
In our study, we demonstrate the first attempt to synthesize an echocardiography video using a diffusion model from a single semantic segmentation map. In order to effectively use the semantic information in the generation process, we proposed spatial adaptive normalization to better incorporate the semantic maps into the denoising model. This results in our model producing more realistic echocardiography videos that are consistent with the input segmentation maps in comparison with previous methods, as we show on the CAMUS dataset. We also examine the shortcomings of recent work and consider potential directions for future investigation.

\newpage
\bibliographystyle{IEEEbib}
\bibliography{bibliography}
\end{document}